\documentclass[prl,reprint,superscriptaddress,nofootinbib]{revtex4-1}
\usepackage{graphicx,amssymb}
\usepackage{bbm}
\usepackage{color}
\usepackage{ulem}

\newcommand{\ben}{\begin{enumerate}}
\newcommand{\een}{\end{enumerate}}
\newcommand{\bit}{\begin{itemize}}
\newcommand{\eit}{\end{itemize}}

\newcommand{\beqa}{\begin{eqnarray}}
\newcommand{\eeqa}{\end{eqnarray}}
\newcommand{\beq}{\begin{equation}}
\newcommand{\eeq}{\end{equation}}
\newcommand{\bay}{\begin{array}}
\newcommand{\eay}{\end{array}}

\def\gsim{\ \rlap{\raise 3pt \hbox{$>$}}{\lower 3pt \hbox{$\sim$}}\ }
\def\lsim{\ \rlap{\raise 3pt \hbox{$<$}}{\lower 3pt \hbox{$\sim$}}\ }

\def\lt{\left}
\def\rt{\right}

\def\lag{{\cal L}}
\def\ipb{{\rm pb}^{-1}}
\def\ifb{{\rm fb}^{-1}}
\def\MET{{E_T\!\!\!\!\!\!\!/\ \ }}
\def\AA{{\cal A}_{1\ell}}

\def\bracket#1#2 {\mathinner{\langle{#1}|{#2}\rangle}}

\definecolor{darkgreen}{rgb}{0,0.6,0}

\begin{document}

%\vspace*{3cm}
\preprint{UCI-HEP-TR-2011-07}

\title{Asymmetric Leptons for Asymmetric Tops}

\author{Arvind Rajaraman}
\affiliation{Department of Physics \& Astronomy, University of California, Irvine, CA 92697}
\author{Ze'ev Surujon}
\affiliation{Department of Physics \& Astronomy, University of California, Riverside, CA 92521}
\author{Tim M.P. Tait}
\affiliation{Department of Physics \& Astronomy, University of California, Irvine, CA 92697}

\vspace*{5mm}

\begin{abstract}
We propose an efficient  method to explore models
which which produce like-sign tops at the LHC, using 
the total charge asymmetry of single lepton events
instead of like-sign dileptons. 
As an example, the method is implemented on a $Z'$ Model,
which can explain the top pair forward-backward asymmetry at Tevatron.
We show that a large region of the parameter space of this model can
be reached using the existing data set at the LHC.
\end{abstract}

\maketitle

%%%%%%%%%%%%%
\section{Introduction}
%%%%%%%%%%%%%

Clarifying the nature of electroweak symmetry breaking (EWSB) is one of the primary
missions of the LHC \cite{Morrissey:2009tf}.  Whatever the agent of
EWSB, it must couple most strongly to the the most massive particles of the Standard Model (SM),
and it is imperative to examine the properties of the heavy quarks and gauge bosons in order
to confirm the SM predictions for their properties.  In particular, the top quark as the
most massive particle discovered to date and the only fermion whose mass lies close to the
electroweak scale itself, is a natural laboratory to explore these questions.  The Tevatron
program has successfully discovered top, measured its mass, and verified many of its expected
features.  As the LHC collects data in earnest, it acts as a top factory and offers unprecedented
potential to examine top quark properties and study production at high energy.

In fact, the Tevatron may already be providing hints for new physics in the top sector.
The observable of primary interest is the forward-backward asymmetry in top production,
\beq
   A^t_{\rm FB}\equiv{N(\Delta y>0)-N(\Delta y<0)\over N(\Delta y>0)+N(\Delta y<0)},
\eeq
where $\Delta y=y_t-y_{\bar t}$ is the difference between the rapidity
of the top and that of the anti-top.
It characterizes how often the top (as opposed to anti-top) tends to go in
the direction of the incoming quark in the reaction $q \bar{q} \rightarrow t \bar{t}$
as observed from the $t\bar t$ center of mass frame.
The measurements \cite{newcdf,cdf,d0,FB-asym}
show an interesting deviation from the expectations of the Standard Model,
where it receives negligible
leading order contributions from electroweak production 
$q \bar{q} \rightarrow Z^* \rightarrow t \bar{t}$,
and small (but in principle measurable) contributions
of  $6\pm1\%$ at next-to-leading order in
QCD \cite{Antunano:2007da}.
A recent update to the measurement \cite{FB-asym} finds
$A^t_{\rm FB}=15.8\pm7.5\%$ and further
indicates that the deviation
is small for top quarks produced with small invariant mass, but grows large 
($48\pm11\%$ measured compared to the SM prediction of $9\pm1\%$)
for invariant masses $M_{t \bar{t}} \geq 450$~GeV.  This feature is exciting because it is 
consistent with the expected influence of
heavy physics operating just beyond the kinematic reach of the Tevatron.
Inspired by these results, a plethora of theoretical proposals have appeared
\cite{Jung:2009jz,Djouadi:2009nb,Rodrigo:2008qe,Jung:2009pi,Shu:2009xf} attempting to
explain it in terms of heavy new physics.

%For $t\bar t$ pairs with $\lt|\Delta y\rt|>1$, it was found to be
%$61\pm26\%$ (instead $\sim12\pm1\%$, which is the SM prediction).

If this measurement does indeed represent
a glimpse of heavy physics just beyond the reach of Tevatron, there is potential for
enormous deviations at the LHC 
\cite{Rodrigo:2010gm,Blum:2011up,Cao:2011ew,AguilarSaavedra:2011vw}, 
whose large center-of-mass energy allows it to rather
easily produce states too heavy for the Tevatron.  The specific signatures at the LHC
can help distinguish between particular models.  For example, some models produce
resonances in $t \bar{t}$ production, whereas others contain new particles decaying into
top and a light quark.  Some models produce {\it like-sign tops} ($tt$ or $\bar t \bar{t}$)
\cite{Cao:2011ew}, and the separate rates of $tt$ and $\bar{t} \bar{t}$ encode information
about the couplings of the new states.  In fact, like-sign top pairs are a striking signal
of physics beyond the Standard Model, one with very little genuine physics background.

The typical signature of like-sign top production uses semi-leptonic decays to measure the
charge of both of the decaying top quarks (along with the $b$-tagged jets
and missing energy such decays produce).  It is a striking signal with very little SM background
(predominantly from jets faking one of the leptons),
but it does come at the cost of requiring both tops to decay into either an electron or muon,
a combined branching ratio of $\left( 2 / 9 \right)^2 \sim 5\%$.  With very limited statistics,
this may severely limit the effectiveness of the signature.

A related observable is the {\it single lepton charge asymmetry}, which
also looks at top pair production, comparing the
number of top decays producing a positive charged lepton with
the number producing a negatively charged lepton,
\beq
   \AA\equiv\frac{N(\mbox{top pair} \to 1\ell^+)-N(\mbox{top pair} \to 1\ell^-)}
   {N(\mbox{top pair} \to 1\ell^+)+N(\mbox{top pair} \to 1\ell^-)}~.
\eeq
Events containing two or more isolated leptons are 
vetoed\footnote{Note also that similar charge asymmetries have been suggested
in the context of single top production~\cite{ChAsym}.}.
 It is aimed at
like-sign top production, since $t \bar{t}$ processes necessarily produce no
asymmetry.  It further specializes to theories where the number of $tt$ pairs is different
from the number of $\bar{t} \bar{t}$ pairs, making use of the fact that the LHC is a
$pp$ collider, with more valence quarks than anti-quarks available in the initial state.
The primary advantage is that it captures a larger fraction of the
like-sign top production, since one top decays hadronically, with a net
branching ratio of $2 (2/9)(2/3) \sim 30\%$.  Thus, with limited statistics it
may be able to show a deviation which would not yet be significant in a traditional
dilepton-based like-sign top reconstruction.  
Even with enough statistics for the traditional like-sign top search, it provides a separate
handle with different systematics to help pin-down the like-sign top signal.
For example, it is less sensitive to the
fake background from a jet faking a lepton, assuming that the charge assigned to the
mis reconstructed lepton is roughly $50\%$ positive and $50\%$ negative.

In this article, we examine the prospects to use the single lepton charge asymmetry
to make an early identification of physics beyond the Standard Model in the form
of an anomalous $tt$ production.  The technique itself is general, but we apply it
in particular to the model of \cite{Jung:2009jz}, which invokes a $Z^\prime$ with
flavor off-diagonal couplings to explain the Tevatron
top forward-backward measurement.  We find the single lepton charge asymmetry
to be a powerful test of such models, and that a significant portion of the parameter space can
be reached with modest amounts of data.

%%%%%%%%%%%%%%%%%%
\section{An Illustrative Model}
%%%%%%%%%%%%%%%%%%

To illustrate the utility of the single lepton charge asymmetry measurement, 
we consider a model
containing a neutral vector $Z^\prime$ whose interactions are given by,
\beq
   \delta\lag=Z^\prime_\mu\bar u_R \gamma^\mu \lt(g_X t_R+g^\prime_X u_R\rt)
   +{\rm c.c.}.
\eeq
It was shown in~\cite{Jung:2009jz} that this model can  generate the observed %$t-\bar{t}$
forward-backward asymmetry 
for viable choices of the parameters; for example
the parameter choices
\beq
   M_{Z^\prime}=160~{\rm GeV}, \,\,\,\alpha_X=0.024,
   \,\,\,\alpha^\prime_X\approx 0.002,
   \label{eq:bestpoint}
\eeq
where $\alpha\equiv g^2/4\pi$, and so on.
As shown in~\cite{Jung:2009jz}, $\alpha'_X$ cannot be much larger
without running into constraints from dijet searches, and we will
ignore it for the purposes of our discussion.
This assumption does have some impact when we discuss
sources of fake events which would contribute to the
$t\bar t$ cross section measurement.  Ref.~\cite{Jung:2009jz}
further focuses on $Z^\prime$s lighter than the top itself, in order to
evade constraints from like-sign top production at the Tevatron.

\section{Top Pair Production Cross Section}

Before examining the single lepton charge asymmetry, we consider
the ramifications of the $Z^\prime$ model on the top pair production rate
at the LHC.  The $Z^\prime$ will contribute to $u\bar u\to t\bar t$, interfering
with the $u \bar{u}$-initiated SM process, and will in addition result in
the processes $u u\to t t$ and $\bar{u} \bar{u} \to \bar{t} \bar{t}$
 (the imbalance of which results in the single lepton charge asymmetry).
Whether these two latter processes contribute to a given measurement of
top pair production depends on the top decay modes under consideration.  
``Dilepton" top pair events typically require that the
 leptons be of opposite charge to suppress fake backgrounds, and will not register
 $tt$ or $\bar{t} \bar{t}$ events.  The ``lepton + jets" mode in which one top decays
 semi-leptonically and the other hadronically, will measure the sum of 
 $t \bar{t} + t t + \bar{t} \bar{t}$ production.  Thus, the $Z^\prime$ model could
 reveal itself either through a discrepancy between the top pair production cross section
 measured in the lepton + jets mode and the SM expectation, or in tension between
 measurements of the lepton + jets mode and the dilepton mode.
 
 At the current time, ATLAS~\cite{ttbar-ATLAS} and CMS ~\cite{ttbar-CMS}
 (which so far has only released dilepton-based measurements)
 measurements of the top pair production rate
 derived from about 2.9 pb$^{-1}$ of integrated luminosity have
 large enough uncertainties on the individual measurement channels so as to make
 it difficult to imagine resolving tension between the dilepton and lepton + jets
 measurements.  However, the ATLAS measurement of
 $142 \pm 60$ pb~\cite{ttbar-ATLAS} in the combined $e + \mu$ single lepton channels
 nevertheless contains useful information.

In order to estimate the rate of $tt$ production
which effectively contributes to the ATLAS measurement, we
begin with a sample of ordinary SM $t\bar t$, generated at the parton level with
Madgraph/Madevent \cite{Alwall:2007st}, with the CTEQ6L1 parton distribution
functions (PDFs) 
\cite{Nadolsky:2008zw} and a renormalization/factorization scale of
$m_t=172.5$~GeV.  We apply a $K$-factor
of $K=1.67$ to match the SM NNLO rate of $164.6^{+11.4}_{-15.7}$~pb
\cite{Moch:2008qy}.
The events are hadronized and showered
by Pythia \cite{Sjostrand:2006za}.  The detector response is simulated with PGS4 
\cite{PGS} using the default  {\tt pgs\_card\_atlas.dat} card file, using the
$k_T$ jet algorithm with cone size $R=0.4$.  
We apply the ATLAS 
$1\mu+\MET+\!\!\ge\!\!4j$~($b$-tagged) selection criteria ~\cite{ttbar-ATLAS}
to find the fraction of events accepted by the ATLAS analysis.

\begin{figure}
  \centering
    \includegraphics[width=0.47\textwidth]{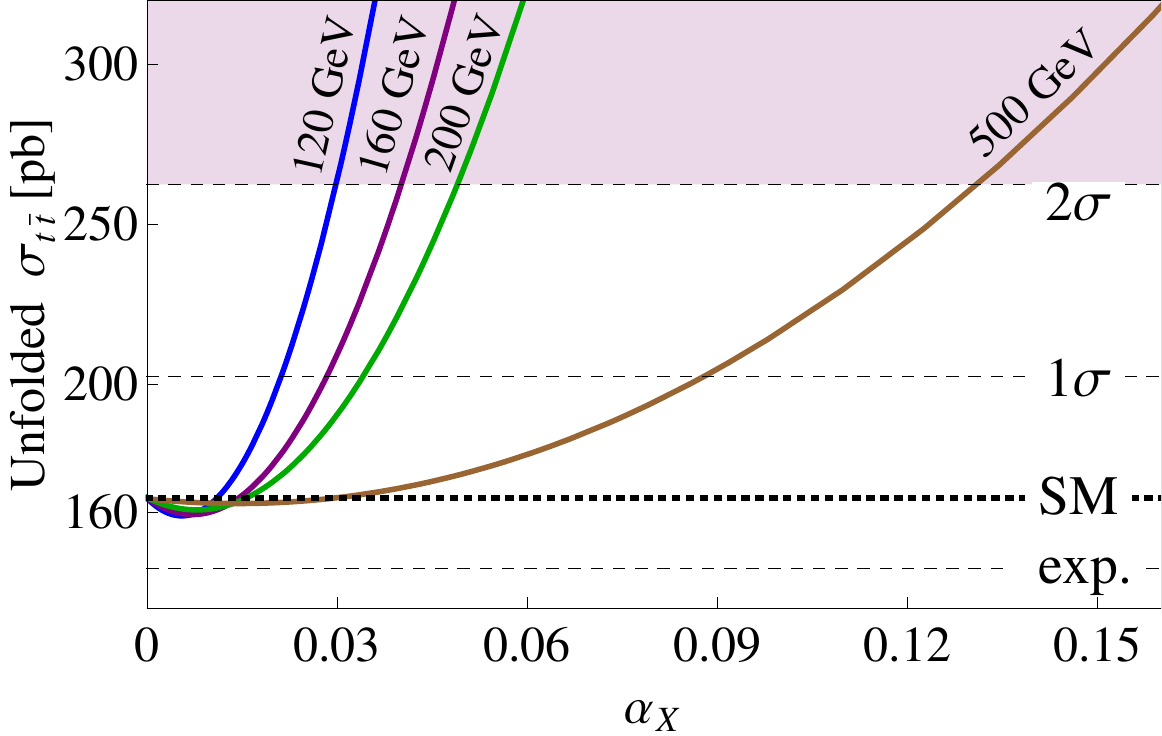}
  \caption{\footnotesize The effective
  $t\bar t$ cross section which would be reported
  by ATLAS ($7$~TeV), as function of $\alpha_X$, for
  several values of the $Z^\prime$ mass.
  The shaded region is $2\sigma$ away from the ATLAS single-lepton
  ($e/\mu$ combined) measurement \protect{\cite{ttbar-ATLAS}}.
  \label{fig:xsec-alpha-LHC}}
\end{figure}

Following the same methodology, we generate $t \bar{t}$ and $tt$ events
 (the rate of $\bar{t}\bar{t}$ is negligible for data sets up to a few fb$^{-1}$)  in the $Z^\prime$
model.  We continue to apply $K=1.67$ for the $t \bar{t}$ rate, but make the
more conservative choice of leaving the $tt$ events at their strict tree level estimate.
We apply the same reconstruction cuts, determining the actual efficiency for $tt$ events
$\varepsilon(tt)$ to pass them, and then unfold with the SM efficiency
$\varepsilon(t\bar t)_{ \rm SM}$.
The effective unfolded cross section to be compared with the ATLAS measurement is thus,
\beq
   \sigma_{\rm unfolded}(t\bar t)= \frac{
   \sigma (t\bar t) \varepsilon(t\bar{t}) +
   \sigma(tt) \varepsilon(tt)}{\varepsilon(t\bar t)_{ \rm SM}}~.
\eeq
In Fig.~\ref{fig:xsec-alpha-LHC}, we show the effective (including both like-sign and
opposite-sign) top pair cross section which would be measured in the
single muon + jets
channel, as a function of $\alpha_X$ and for several choices of $M_{Z^\prime}$.  
Overlaid the predictions of the $Z^\prime$ model 
is the ATLAS measurement and its one and two sigma uncertainty bands.  We see that for small
$Z^\prime$ masses, ATLAS is already probing some of the relevant parameter space
to explain the $A_{\rm FB}^t$ measurement through its $t \bar{t}$ cross section measurement, but
much of the parameter space remains unconstrained.

%%%%%%%%%%%%
%%%%%%%%%%%%

%%%%%%%%%%%%%%%%%%%%
\section{Single Lepton Charge Asymmetry}
%%%%%%%%%%%%%%%%%%%%

We now turn to the single-lepton charge asymmetry $\AA$ in the context of
the $Z^\prime$ model. We are looking for the excess leptons from top
decay in the process $u u\to t t$,
driven by the large up quark valence PDFs.
there will be an excess of positively charged leptons, resulting in
a non-zero value of 
$\Delta N=N(\ell^+)-N(\ell^-)$ in the $\ell+4j$~($b$-tagged) sample.
We  will find that measuring this excess is an effective way to probe this model
even at the early LHC.

One of the key issues is understanding the non-top pair backgrounds, some of which
themselves exhibit a single lepton charge asymmetry.  In many cases, these
backgrounds can be estimated with the
help of the data itself.
We focus on the single muon plus at least four jets with at least one jet $b$-tagged
($\mu~+ \ge\!\!4j$~($b$-tagged))
for our analysis, because it is clean, with
the smallest expected ``fake" (non-top) background~\cite{ttbar-ATLAS}.
Apart from being the cleanest, it also 
avoids contributions from $t Z^\prime$ events (the expected rate of
which depends on $\alpha^\prime_X$).
Such contributions do not typically
pass the $\ell+\ge\!\!4j$ cuts, though they would be present in the
$\ell+3j$ sample.

The $\mu~+ \ge\!\!4j$~($b$-tagged) channel is defined by the following
selection criteria~\cite{ttbar-ATLAS}:
\bit
   \item The event contains exactly one muon:
   \bit
      \item $p_T>20$~GeV and $|\eta|<2.5$;
      \item separation $\Delta R>0.4$ from any jet with $p_T>20$~GeV;
      \item scalar sum of transverse momenta for all tracks within a cone of $\Delta R=0.4$
         is less than $4$~GeV;
   \eit
   \item $\MET>20$~GeV and $\MET+m_T(W)>60$~GeV,
      where $m_T(W)\equiv\sqrt{2p_T(\mu)\MET\lt\{1-\cos\lt[\phi(\mu)-\phi(\MET)\rt]\rt\}}$;
   \item at least four jets with $p_T>25$~GeV and $|\eta|<2.5$;
   \item at least one $b$-tagged jet.
\eit

The background to our signal comes from many processes which produce a
single muon along with $\geq 4$ jets, passing these cuts.
Some of these such as $t \bar{t}$ itself, $Z$ + jets and 
QCD jet production (where a jet is mis-tagged as a lepton) do not possess an intrinsic lepton charge
asymmetry, but for a finite dataset may fluctuate, and thus represent an important
uncertainty on a measured value.  Others, such as $W$ + jets or single top production
\cite{Dawson:1984gx} contribute directly to the asymmetry, and must be properly accounted for
in order to isolate the contribution from top pair production.  The full set of backgrounds considered
is listed in Table~\ref{tab:bkg}.
For each background process, we determine the ``effective cross section", defined as the cross section
after imposing the ATLAS $\mu~+ \ge\!\!4j$~($b$-tagged) cuts.  The ATLAS cross section measurement provides
estimates for the rates derived from Monte Carlo and passed through the full ATLAS detector 
simulation~\cite{ttbar-ATLAS}.
We list the total effective cross sections for the background processes
in the second column of Table~\ref{tab:bkg}.

The $W$ + jets and single top processes also result in their own non-zero contributions to $\AA$
driven by the fact that at large parton $x$ there are roughly twice as many $u$
as $d$ valence quarks in the proton, resulting in $\AA \sim 1/3$.  In practice, both processes also
receive contributions from subprocesses initiated by sea quarks and gluons, 
which are charge-symmetric and lead to $\AA < 1/3$.  
Since these are subdominant components of the background anyway, we make the
conservative choice to set them both to $1/3$ in the current study.  For example, the charge 
asymmetry
in inclusive $W \to \mu \nu$ events has been measured to range from $0.15$ to $0.3$ for muon 
rapidities between 
0 and 2.0 \cite{ChAsym-ATLAS}.  For more precise studies, it would be desirable to determine 
these contributions more precisely, either from Monte Carlo estimates, or directly from the
data (using larger datasets).  With precise enough independent determinations, these
background biases may be subtracted out
to obtain a clean measurement of the excess over the SM.
But even if the asymmetries were known precisely, the systematic uncertainties
on the effective cross-sections of the background processes must be propagated.
%
%%%%%%%%%%%%%%%%%%%%%%%%%%%%
\begin{table}
   \centering
   \begin{tabular}{ccc}
      \hline
      Background Process & $\sigma_{\rm eff}[{\rm pb}]$ & $\AA$\\
      \hline\hline
      $t\bar t$ & $5.17\pm1.17$ & $0$\\
      \hline
      $W+$jets & $0.586\pm0.552$ & $+1/3$\\
      \hline
      $Z+$jets & $0.034\pm0.034$ & $0$\\
      \hline
      Single top & $0.241\pm0.069$ & $+1/3$\\
      \hline
      QCD jets & $0.276\pm0.173$ & $0$\\
      \hline\hline
      SM combined & $6.31\pm1.31$ & $+0.044$\\
      \hline
   \end{tabular}
   \caption{\footnotesize An estimate of the SM background
   for the single muon charge asymmetry $\AA$.
   The corresponding effective cross-sections
   %through the $\mu+\ge\!\!4j$ cuts (
   inferred from~\cite{ttbar-ATLAS}
   are also indicated, along with their systematic errors (including the luminosity error), 
   as well as our
   conservative choices for their intrinsic single lepton charge asymmetry.
   \label{tab:bkg}}
\end{table} 
%%%%%%%%%%%%%%%%%%%%%%%%%%%%
The resulting systematic error on the difference between
single positive and single negative lepton events ($\Delta N$) is
%
%from the asymmetric $W$ + jets and single top
%processes can be estimated from Table~\ref{tab:bkg} as,
%
\beqa
   \delta_{\rm sys}(\Delta N) &=&
   \frac{1}{3}\sqrt{\delta\sigma_{\rm eff}^2(W+{\rm jets})
   +\delta\sigma_{\rm eff}^2({\rm single\ top})}\times L\nonumber\\
   & \simeq &0.185 \lt(\frac{L}{\ipb}\rt),
\eeqa
where $L$ is the collected integrated luminosity.
In addition, all processes
contribute to the statistical error on $\Delta N$,
\beq
   \delta_{\rm stat}(\Delta N)=\sqrt{N_{SM}}
   =\sqrt{6.31\lt(\frac{L}{\ipb}\rt)}.
\eeq
Note that the small effective cross sections of $W+$jets and single top events
ensure that their systematic effect is small even if our estimate of the
corresponding asymmetries is bad. 
This is of course based on the assumption that the main background
process, namely $t\bar t$, has little or no asymmetry.

For the $Z^\prime$ model, we simulate the rates of $tt$ production as
described above, and select events in the $\mu+\ge\!\!4j$~($b$-tagged)
channel.  This results in the prediction for $\AA$ corresponding to a given
choice of $m_{Z^\prime}$ and $\alpha_X$.  
The significance of the measurement is,
\beq
   s=\frac{\sigma_{\rm eff}[tt]\lt(\alpha_X,m_{Z'}\rt)}
   {\sqrt{\delta_{\rm stat.}^2(\Delta N)+\delta_{\rm sys.}^2(\Delta N)}}.
\eeq

In Fig.~\ref{fig:asym-reach}, we plot the contours of
fixed expected significance of the $\AA$ measurement
in the $(m_{Z'},\alpha_X)$ plane, both for the current data
corresponding to $35\ipb$, as well as for a future measurement with
1~fb$^{-1}$.  Already, the current data allows one to constrain
a significant portion of the parameter space consistent with the Tevatron
measurement, if the central value of $\AA$ turns out to be zero.  With
$50~\ipb$ of data, the parameter choice
($M_{Z^\prime}=160$~GeV and $\alpha_X=0.024$) mentioned above
can be excluded at better than the $95\%$ CL
if the SM expectation
is obtained.  With 1~$\ifb$ of data, the same point can be 
discovered at the $5\sigma$
level.
%
%%%%%%%%%%%%
\begin{figure}
  \centering
    \includegraphics[width=0.47\textwidth]{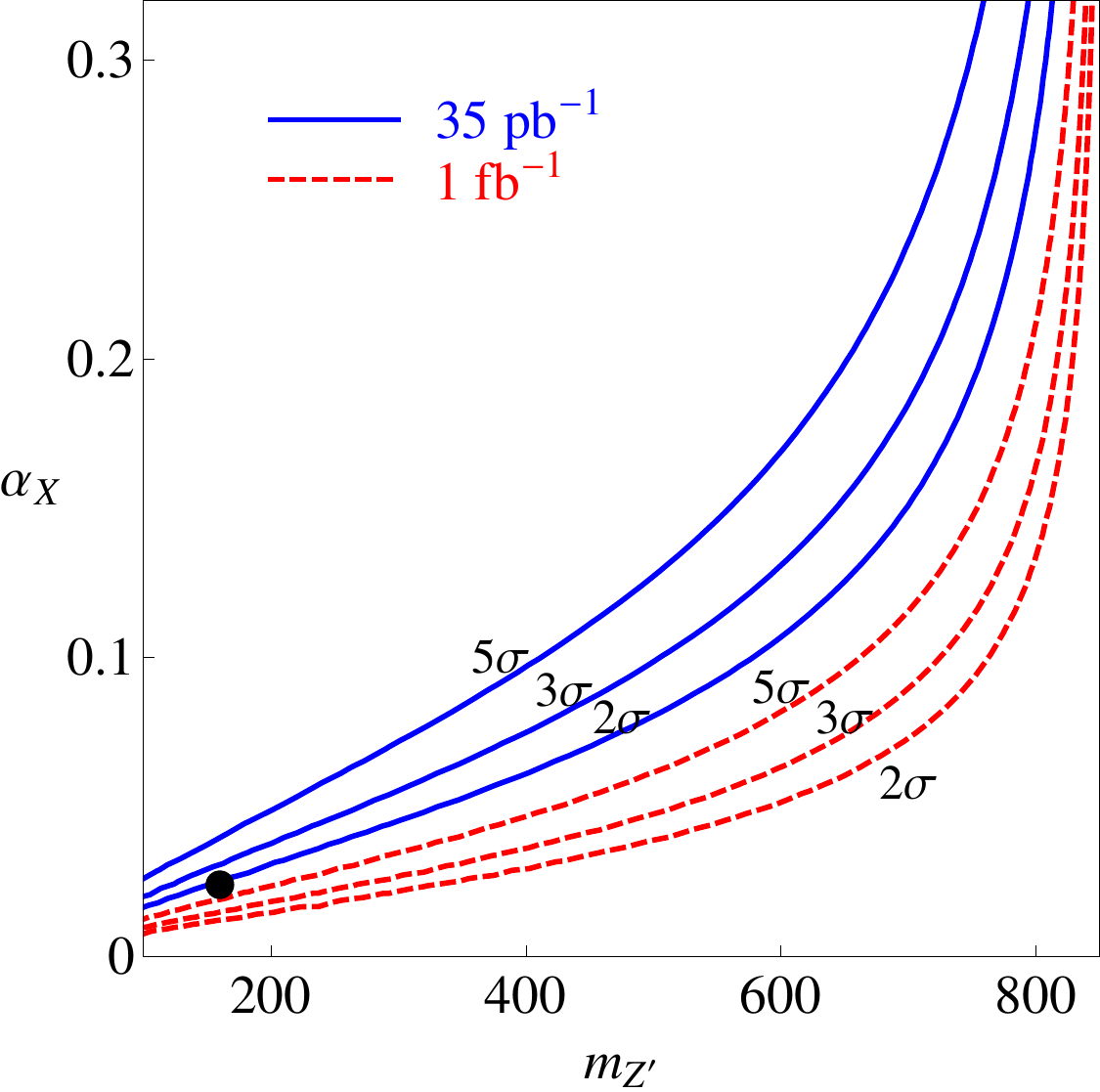}
    \caption{\footnotesize Significance levels ($2\sigma$, $3\sigma$, $5\sigma$)
    for $L=35~\ipb$ and $L=1~\ifb$ in the $\lt(m_{Z'},\alpha_X\rt)$
    plane, from measurement of the single-lepton charge asymmetry,
    using the $1\mu+\MET+\!\!\ge\!\!\!4j$~($b$-tagged) channel at $7$~TeV.
    The black dot represents the point
    $\lt(m_{Z'}=160~{\rm GeV},\ \alpha_X=0.024\rt)$
    discussed in the text.
    \label{fig:asym-reach}}
\end{figure}
%%%%%%%%%%%%

%%%%%%%%%%%%
%\begin{figure}
%  \centering
%    \includegraphics[width=0.46\textwidth]{plot-xsec-vs-mZ.pdf}
%    \caption{\footnotesize Effective $tt$ cross-section through the ATLAS
%    cuts of $\mu+\MET+\ge\!\!4j$ tagged ($7$~TeV),
%    as function of the $Z'$ and for $\alpha_X=0.024$.
%    %The fitted curve is a second-order polynomial in $1/m_{Z'}$.
%    \label{fig:tt-vs-mZ}}
%\end{figure}
%%%%%%%%%%%%

%%%%%%%%%%%%%%%%%%%
\section{Conclusions}
%%%%%%%%%%%%%%%%%%%
We have discussed a simple method to probe models which produce
like sign tops at the LHC, through observation of an imbalance between the
number of top pairs leading to a single positively charged lepton and the
number of negatively charged single leptons.  It has the advantage of requiring
less statistics than a traditional like-sign top search relying on two like-sign leptons
from the top decays.  Even when statistics are sufficient for
the like-sign top measurement to be effective, it remains a good complement 
to such a measurement,
because it is sensitive to different backgrounds and different experimental errors.

We have illustrated how it
works for a particular $Z^\prime$ model designed to explain the Tevatron
measurements of $A_{\rm FB}^t$, and find that it is expected to be able to say
something non-trivial even with the current data set of $35~\ipb$.  With larger
data sets it can exclude or discover a significant portion of the relevant parameter
space.  Nonetheless, it is much more general, and can
be applied to any model producing like-sign tops.

\acknowledgements
We would like to thank Daniel Whiteson for discussions.
ZS thanks Subhaditya Bhattacharya, Will Shepherd and Shufang Su
for helpful discussions.
The work of AR is
supported in part by NSF Grants PHY-0653656 and PHY-0709742.
ZS is supported in part by DOE grant FG03-94ER40837.
TMPT is supported by NSF grant PHY-0970171 and in addition he gratefully acknowledges the hospitality of the SLAC theory group, where part of this work was completed.

\end{document}